\documentclass[12pt]{article}
\usepackage{amsmath,amsthm,amscd,amssymb}
\usepackage{latexsym}
\usepackage{epsf}
\usepackage{epsfig}
\begin{document}

\title{Acceleration-Induced Nonlocality:\\ Uniqueness of the Kernel}
\author{C. Chicone\\Department of Mathematics\\University of 
Missouri-Columbia\\Columbia, Missouri 65211, USA \and B. 
Mashhoon\thanks{Corresponding author. E-mail:
mashhoonb@missouri.edu (B. Mashhoon).} \\Department of Physics and
Astronomy\\University of Missouri-Columbia\\Columbia, Missouri 65211, USA}
\maketitle

\maketitle

\begin{abstract}
  We consider the problem of uniqueness of the kernel in the nonlocal 
  theory of accelerated observers. In a recent work \cite{14}, we 
  showed that the convolution kernel is ruled out as it can lead to 
  divergences for nonuniform accelerated motion. Here we determine the 
  general form of bounded continuous kernels and use observational 
  data regarding spin-rotation coupling to argue that the kinetic 
  kernel given by $K(\tau ,\tau')=k(\tau')$ is the only physically 
  acceptable solution.  
\end{abstract}
\section{Introduction}

The special theory of relativity deals with physics in Minkowski spacetime \cite{1}. The test observers in this theory are in general noninertial; for the results
of measurements of a noninertial observer, the theory asserts that such an observer is locally inertial (``Hypothesis of Locality"). In this way, Lorentz invariance
may be applied in a pointwise manner to make physical predictions regarding what accelerated observers measure. Thus Lorentz invariance and the hypothesis of
locality together constitute the pillars of the special theory of relativity.

The inhomogeneous Lorentz transformations 
(i.e. elements of the Poincar\'e group) 
connect the physical measurements of ideal inertial observers. 
These have a special significance: 
The fundamental laws of microphysics involve quantities that 
are ultimately measured by such ideal inertial observers. 
However, all actual observers are accelerated. 
It is therefore necessary to specify how the 
measurements of an accelerated observer 
can be connected with the basic laws of physics;
that is, a connection is needed between the accelerated and 
inertial observers. 
The hypothesis of locality provides such a 
connection as it postulates that an accelerated observer is at each 
instant equivalent---in terms of
measurements of physical quantities---to an otherwise identical 
hypothetical inertial observer that has the same state 
(i.e. position and velocity) as the
accelerated observer. The approximate nature of this assumption, 
which is natural from the standpoint of Newtonian mechanics, 
was discussed by Lorentz in the
specific context of his electron theory \cite{2}. 
The hypothesis of locality extends to all measuring devices the assumption 
that the rods and clocks of the
standard theory of relativity are locally inertial \cite[p. 60]{1}.

If the duration of the basic phenomenon measured by the noninertial 
observer is such that its velocity vector and spatial reference frame
do not change appreciably 
over this time interval,
then the observer may be regarded as inertial and the hypothesis of locality 
is valid; that is, the acceleration 
of the observer is locally immaterial. To quantify
this criterion, 
we note that there are certain invariant 
acceleration time scales $\mathcal{L}/c$, 
given typically by $c/g$ and $1/\Omega$, associated with a
noninertial observer. 
These are related to the magnitudes of the observer's translational 
acceleration $\mathbf{g}$ and the rotational frequency
\boldmath$\Omega$\unboldmath \; of its spatial frame, respectively \cite{3,4}. If $\lambda/c$ is the intrinsic time scale of the phenomenon under observation, then
the expected deviation from the 
hypothesis of locality is $\sim \lambda/\mathcal{L}$. 
Such a deviation turns out to be too small to be detectable in most physics
experiments. To illustrate this point, let us note that experiments are 
typically performed in a laboratory fixed on the rotating Earth, 
where $c/g_\oplus \sim 1$
year and $1/\Omega_\oplus \sim 4$ hours, 
while for a laser beam $\lambda/c\sim 10^{-15}$ second. 
In this way, one can account for the great success of the standard
theory of relativity. As a matter of principle, however, 
it would be interesting to construct a viable theory of accelerated 
observers in Minkowski spacetime that
goes beyond the hypothesis of locality. Such a nonlocal theory of 
accelerated systems is described in section~2. 
For the sake of concreteness, electromagnetic
radiation fields are 
considered throughout this paper; 
however, the final results in their general form would be valid for any field.

The nonlocal theory of section~2 involves a kernel that needs to be 
determined on the basis of physical principles discussed in section~3.
In this way, a class of bounded continuous kernels is identified. 
In section~4, the resulting
nonlocal theory is confronted with observational data regarding the 
measurement of electromagnetic radiation fields by a uniformly 
rotating observer. Thereby a
unique kernel is tentatively identified. 
The final section contains a discussion of our results.

\section{Nonlocality of accelerated observers}

We consider a global inertial frame in Minkowski spacetime with coordinates 
$x^\alpha=(ct,\mathbf{x})$. 
This is the only coordinate system that is needed here; in
particular, we avoid the use of ``accelerated coordinate systems" 
due to their fundamental limitations associated with the measurement of 
distance in such systems
\cite{3,4}.

The accelerated observer follows a worldline with tangent vector 
$\lambda^\mu_{\;\;(0)}=dx^\mu /d\tau$, where $\tau$ 
is the proper time along the path. The observer
refers its measurements to an orthonormal tetrad frame 
$\lambda^\mu_{\;\;(\alpha)}$ defined along its worldline such that
\begin{equation}\label{eq1} 
\frac{d\lambda^\mu_{\;\;(\alpha)}}{d\tau}
=\phi_\alpha ^{\;\;\beta}\lambda^\mu _{\;\;(\beta)}.
\end{equation}
Here the scalars $\phi_{\alpha \beta}$ form an antisymmetric acceleration 
tensor such that the ``electric" and ``magnetic" parts correspond to the 
acceleration
$\mathbf{g}$ of the observer $(\phi_{0i}=g_i/c)$ and the rotation frequency 
\boldmath$\Omega$\unboldmath\; of its spatial frame
$(\phi_{ij}=\epsilon_{ijk}\Omega_k)$, respectively. We note that along its 
worldline the accelerated observer passes through a continuous infinity of 
hypothetical
instantaneously comoving inertial observers each with the instantaneous 
tetrad frame $\lambda^\mu_{\;\;(\alpha)}$. 
To avoid unphysical situations involving the
expenditure of an infinite amount of energy in order to keep the observer 
accelerated, we assume that the acceleration is turned on at $\tau_0$ 
and turned off at a later time $\tau_1$.

Let $f_{\mu \nu}$ be an electromagnetic radiation field in Minkowski spacetime; in fact,
$f_{\mu \nu}$ is the Faraday tensor as measured by the standard set of
static inertial observers in the background global frame. 
Let $F_{\alpha \beta}$ be the corresponding radiation field 
as measured by the accelerated observer. The
hypothesis of locality implies that the field as measured by the accelerated observer is given at each point along the worldline by the field as measured by the
hypothetical momentarily comoving inertial observer. For such an observer the measured field is the 
projection of $f_{\mu\nu}$ upon its tetrad frame by Lorentz
invariance, i.e.
\begin{equation}\label{eq2}
 \hat{f}_{\alpha \beta}
=f_{\mu \nu}\lambda^\mu_{\;\;(\alpha)}\lambda^\nu_{\;\;(\beta)}.
\end{equation}
The accelerated observer passes through an infinite sequence of such momentarily comoving inertial observers; therefore, the most general linear relationship
between $F_{\alpha \beta}$ and $\hat{f}_{\alpha \beta}$ consistent with 
causality is \cite{5}
\begin{equation}\label{eq3} 
F_{\alpha \beta}(\tau)=\hat{f}_{\alpha \beta}(\tau)
+\int^\tau_{\tau_0}K_{\alpha\beta}^{\;\;\;\;\gamma
\delta}(\tau,\tau')\hat{f}_{\gamma\delta}(\tau')d\tau'.\end{equation}
It is important to recognize that this ansatz only involves spacetime scalars; 
moreover, the kernel $K$ must vanish for an inertial observer. Therefore, the
nonlocal ansatz \eqref{eq3} is physically reasonable if $K$ is 
related to the acceleration of the observer.

Equation~\eqref{eq3} has the form of a Volterra integral equation of the second kind \cite{6}. It follows from Volterra's theorem that in the space of continuous
functions the relationship between $F_{\alpha \beta}$ and $f_{\mu \nu}$ is unique \cite{6}. This uniqueness result has been extended to the Hilbert space of
square-integrable functions by Tricomi~\cite{7}. 
In this sense, therefore, we assume boundedness as well as continuity 
throughout this paper. Further details regarding acceleration-induced
nonlocality can be found in~\cite{5,8}.

\section{Determination of the kernel}

The hypothesis of locality has a consequence that goes against the spirit of 
relativity theory: a pure radiation field can stand 
completely still with respect to a
uniformly rotating observer. This is most easily seen for the case of an 
observer rotating uniformly with frequency $\Omega_0\mathbf{n}$ 
about the direction of
propagation---characterized by the unit
 vector $\mathbf{n}$---of a  plane electromagnetic wave of frequency $\omega$. 
The Fourier analysis of $\hat{f}_{\alpha
\beta}$ in this case reveals that 
$\hat{\omega}=\gamma (\omega\mp \Omega_0)$, where the upper (lower) sign refers to positive (negative) helicity incident
radiation. This result differs from the transverse Doppler effect 
$\gamma\omega$ that involves the time dilation factor 
$\gamma=dt/d\tau$; moreover, the
subtraction and addition of frequencies has a simple intuitive interpretation. 
The electromagnetic radiation field rotates about the direction of 
propagation with frequency
$\omega\; (-\omega)$ for a positive (negative) helicity wave; therefore,
the rotating observer perceives radiation of definite helicity but with frequency
$\omega-\Omega_0\;(\omega+\Omega_0)$. The deviation of this result from the transverse Doppler effect provides an instance of the general phenomenon of spin-rotation
coupling. Partial observational evidence for this general coupling is reviewed in \cite{9}. In the case of electromagnetic radiation with $\omega \gg \Omega_0$,
experimental results in favor of this coupling are available in the microwave and optical domains \cite{9}; moreover, this effect has 
been observed for radio waves
$(\nu \sim 1\,\mbox{GHz})$ as a phase wrap-up in the GPS system as described in \cite{10}. In all the experimentally viable cases at present $\Omega_0/\omega \ll 1$;
in this regime, the spin-rotation coupling has therefore a solid
observational basis for electromagnetic radiation \cite{9,10}.
On the other hand, $\omega '=0$ for positive helicity
radiation of frequency $\omega=\Omega_0$, 
i.e. the wave stands completely still with respect to the observer.
More generally, for oblique incidence $\hat\omega=\gamma(\omega-M\Omega_0)$,
where $M=0,\pm 1,\pm 2,\ldots$, is the multipole parameter such that
$\hbar M$ is the component of the total angular momentum of the radiation
field along the axis of rotation of the observer; as before, a multipole
radiation field with $M=\omega/\Omega_0$ can stand completely still with
respect to the rotating observer~\cite{8,9}. In the case of inertial observers, this
cannot occur since the speed of the observer is always less than the speed of light \cite{11}. Thus in the formula for the Doppler effect $\omega '=\gamma \omega
(1-\mathbf{n}\cdot \mathbf{v}/c)$, $\omega '=0$ implies that $\omega =0$. 
We demand that the same should happen for the accelerated observer 
in the nonlocal theory. 
That is, we postulate that a fundamental radiation field can never stand completely still with respect to {\em any} observer. Thus if $F_{\alpha \beta}$
turns out to be constant in equation~\eqref{eq3}, then $f_{\mu \nu}$ should be constant as well. Expressing the Faraday tensor as a six-vector $f_{\mu\nu}\to
(\mathbf{E},\mathbf{B})$ and writing equation~\eqref{eq2} in matrix notation as $\hat{f}=\Lambda f$, we note that equation~\eqref{eq3} can be written as
\begin{equation}\label{eq4}
 F(\tau )
=\Lambda (\tau)f(\tau )+\int^\tau_{\tau_0}
 K(\tau ,\tau')\Lambda (\tau ')f(\tau ')\,d\tau ',
\end{equation}
where $f(\tau)$ is the restriction of the field measured by the standard static inertial observers to the worldline of the accelerated observer. Thus in
equation~\eqref{eq4} a constant $F$ would imply a constant $f$ only if
\begin{equation}\label{eq5} 
\Lambda_0
=\Lambda (\tau)+\int^\tau_{\tau_0}K(\tau ,\tau')\Lambda (\tau')\,d\tau',
\end{equation}
where $\Lambda_0=\Lambda (\tau_0)$. Once the kernel is determined using equation~\eqref{eq5}, it follows from the Volterra-Tricomi uniqueness theorem that for any
true radiation field $f_{\mu \nu}$, 
the accelerated observer will never measure a constant field
(cf.~\cite{8,9} and the references therein).

To determine the kernel $K(\tau, \tau')$, equation~\eqref{eq5} must be solved under the requirements that (i) $K(\tau ,\tau')$ exists due 
to the temporal variation
of $\Lambda$, so that $K(\tau ,\tau')=0$ if $\Lambda$ is constant, since the kernel must vanish for an inertial observer and (ii) the nonlocal contribution to the
field in equation~\eqref{eq4} is always bounded. 
This latter requirement turns out to be crucial. To see this, let us consider two possible solutions of
equation~\eqref{eq5} assuming that $K(\tau ,\tau')$ is a function of only one variable: $K(\tau ,\tau')=k(\tau ')$ and $K(\tau ,\tau ')=\tilde{k}(\tau -\tau')$. In
the first case, equation~\eqref{eq5} is solved by simple differentiation and the result is \cite{12,13}
\begin{equation}\label{eq6} k(\tau )=-\frac{d\Lambda (\tau )}{d\tau }\Lambda^{-1}(\tau ).\end{equation}
The second case involving a convolution kernel is more complicated; nevertheless, equation~\eqref{eq5} is sufficient to determine $\tilde{k}(\tau -\tau ')$
uniquely. Requirement (i) is satisfied in either case; moreover, they give the {\em same} kernel for {\em uniform} accelerated motion. However, for nonuniform
acceleration the convolution kernel can lead to divergence \cite{14} in contrast to equation~\eqref{eq6}. Thus the first case, where the kernel \eqref{eq6} is
directly proportional to the acceleration of the observer gives an acceptable solution called the {\em kinetic kernel} \cite{14}.

Once an acceptable solution of equation~\eqref{eq5} is available, such as the kinetic kernel $k(\tau)$, then the general solution may be written as
\begin{equation}\label{eq7} K(\tau ,\tau')=k(\tau ')+L(\tau ,\tau')\Lambda^{-1}(\tau '),\end{equation}
where $L$ is a $6\times 6$ matrix that vanishes for constant $\Lambda$ and involves bounded 
continuous functions such that
\begin{equation}\label{eq8} 
\int^\tau_{\tau_0}L(\tau ,\tau')\,d\tau '=0.
\end{equation}
It follows from an application of the general theory of 
Fourier series that over 
the interval $[\tau_0,\tau]$, $L(\tau ,\tau')$ is a linear 
superposition of
functions of the form
\begin{equation}\label{eq9} 
a_k(\tau-\tau_0)\,e^{2\pi ik\frac{\tau '-\tau_0}{\tau -\tau_0}}
\end{equation}
for any integer $k\neq 0$, where $a_k$ are bounded continuous 
matrix-valued functions. In this way, equation~\eqref{eq8} 
is satisfied and all that remains is to
ensure that the time dependence of $a_k(\tau-\tau_0)$ is {\em solely} 
due to the variation of $\Lambda (\tau )$.

The requirement that $L(\tau ,\tau')$ must vanish for a constant $\Lambda$ 
is satisfied by expressing $a_k(\tau -\tau_0)$ as a double series
\begin{equation}\label{eq10} 
a_k=\sum^\infty_{n=1}\sum^\infty_{m=1}C^n_{km }
\left(\frac{d^m\Lambda}{d\tau^m}\right)^n+Q_k,
\end{equation}
where $C^n_{km}$ are constants such that the double series is 
absolutely convergent. Here  $Q_k=Q_k(\Lambda)$ is a $6\times 6$ 
matrix that is constant with respect to $\tau-\tau_0$ and such that
$Q_k(\Lambda)=0$ whenever $\Lambda$ is constant. A  
class of such functions is given by 
\begin{equation}\label{eq11}
 Q_k=\int^\infty_{\tau_0}\Phi_k (\tau')\mathcal{Q}_k(\Lambda (\tau '))\,d\tau',
\end{equation}
where $\Phi_k$ and $\mathcal{Q}_k$ are bounded continuous functions and 
\begin{equation}\label{eq12}
 \int^\infty_{\tau_0}\Phi_k (\tau ')\,d\tau '=0.
\end{equation}
Let us note that $C^n_{km}$ may also be proportional to 
constants of the form given by equation~\eqref{eq11}. 
Combining these results, we may therefore express $L(\tau ,\tau')$ 
as a uniformly convergent series
\begin{equation}\label{eq13}
 L(\tau ,\tau')=\mbox{Re}\, 
\sum_{k\neq 0}a_k(\tau -\tau_0)\,e^{2\pi ik\frac{\tau'-\tau_0}{\tau-\tau_0}},
 \end{equation}
where $a_k$ is given by equations~\eqref{eq10}--\eqref{eq12}. 
Substituting equation~\eqref{eq13} in equation~\eqref{eq7},
 we find the general form of the kernel
$K(\tau ,\tau')$ that satisfies our physical requirements.

We must next consider the physical consequences of the general kernel 
for the nonlocal theory. The field measured by the accelerated observer is obtained from the
substitution of equation~\eqref{eq7} in equation~\eqref{eq4}. We observe that in the calculation of $F(\tau)$ beyond the kinetic kernel, all terms would involve
expressions of the form
\begin{equation}\label{eq14}
{f}_k(\tau -\tau_0)
=\int^\tau_{\tau_0}e^{2\pi ik\frac{\tau'-\tau_0}{\tau-\tau_0}}
f(\tau')\,d\tau',
\end{equation}
for $k\neq 0$. It follows that
\begin{align}\label{eq15}
F(\tau )&=\hat{f}(\tau )
+\int^\tau_{\tau_0}k(\tau ')\hat{f}(\tau ')\,d\tau'
+\mathcal{F}(\tau -\tau_0),\\
\intertext{where}
\label{eq16}
 \mathcal{F}&=\mbox{Re}\, \sum_{k\neq 0}a_k {f}_k .
\end{align}
Thus the measured field consists of a superposition of what 
would be expected on the basis of the kinetic kernel alone together with 
$\mathcal{F}$ that consists of
the extra terms proportional to ${f}_k$. It would be interesting to 
examine the physical consequences of the presence of the extra terms in the 
field as
measured by an accelerated observer. This is done in the next section for 
an observer rotating uniformly with $\Omega_0\ll \omega$, since excellent 
observational
data are available in this case for the coupling of the angular 
momentum of the electromagnetic radiation field to the rotation of the 
observer \cite{9,10}.

\section{Spin-rotation coupling}

The observational results regarding spin-rotation 
coupling for electromagnetic radiation described in \cite{9,10} 
may be used as evidence against the presence of
$\mathcal{F}$ in equation~\eqref{eq15}. 
To this end, we consider a plane monochromatic wave of frequency $\omega$ 
and definite helicity that is normally incident
on an observer rotating uniformly with frequency $\Omega _0\ll \omega$
 on a circle of radius $r$ in the $(x,y)$-plane. 
Experiments indicate that the measured
frequency is $\hat{\omega}=\gamma (\omega \mp \Omega_0)$, 
where the upper (lower) sign refers to incident positive (negative) helicity
radiation
and $\gamma $ is the 
Lorentz factor
corresponding to $\beta=v/c$ with $v=r\Omega_0\ll c$. 
We will compare and contrast this result with the predicted spectrum based 
on equations~\eqref{eq15} and \eqref{eq16}.

The uniformly rotating observer has been discussed in detail in \cite{8,14}.
We assume that for $\tau<\tau_0$ the observer moves along a straight
line with uniform speed ($\beta\ll 1$)  such that $x=r$ and 
$y=\gamma v(\tau-\tau_0)$ and at $\tau=\tau_0$ begins uniform circular
motion with $x=r\cos\varphi$ and $y=r\sin\varphi$, where
$\varphi=\gamma\Omega_0(\tau-\tau_0)$.
The natural orthonormal tetrad frame of the uniformly rotating observer is 
given by
\begin{equation}\begin{split}\label{eq17} 
\lambda^\mu_{\;\;(0)}&=\gamma (1,-\beta\sin \varphi ,\beta\cos \varphi ,0),\\
\lambda^\mu_{\;\;(1)}&=(0,\cos \varphi ,\sin \varphi ,0),\\
\lambda^\mu_{\;\;(2)}&=\gamma(\beta,-\sin \varphi,\cos \varphi ,0),\\
\lambda^\mu_{\;\;(3)}&=(0,0,0,1).\end{split}\end{equation} 
The acceleration tensor $\phi_{\alpha \beta}$ 
is given in this case by a centripetal acceleration
$\mathbf{g}=-\gamma^2v\Omega_0(1,0,0)$ and a rotation frequency 
\boldmath$\Omega$\unboldmath$=\gamma^2\Omega_0(0,0,1)$ with respect 
to the tetrad frame
\eqref{eq17}.

The incident field along the worldline of the observer may be expressed as
\begin{equation}\label{eq18} 
f(\tau )=\frac{1}{2}i\omega A
\begin{bmatrix} \mathbf{e}_{\pm}\\\mathbf{b}_{\pm}\end{bmatrix} 
e^{-i\,\gamma \omega(\tau-\tau_0)}+c.\;c.,
\end{equation}
where $A$ is a constant complex amplitude, ``c.\ c.'' indicates
the corresponding complex conjugate term, 
$\mathbf{e}_\pm=(\mathbf{e}_1\pm i\mathbf{e}_2)/\sqrt{2}$, 
$\mathbf{b}_\pm =\mp i\mathbf{e}_\pm$ and the upper (lower)
sign indicates positive (negative) helicity radiation. Here $\mathbf{e}_1$ and $\mathbf{e}_2$ indicate unit vectors along the positive $x$ and $y$ axes,
respectively. The field as measured by the hypothetical 
comoving inertial observers is given by $\hat{f}=\Lambda f$, where
\begin{equation}\label{eq19} 
\Lambda =
\begin{bmatrix}\Lambda_1 & \Lambda_2\\-\Lambda_2 & \Lambda_1
\end{bmatrix}
\end{equation}
and
\begin{equation}
\label{eq20} \Lambda_1=
\begin{bmatrix}
\gamma \cos \varphi & \gamma \sin \varphi & 
  0\\-\sin \varphi &\cos \varphi & 0\\ 0 & 0 &\gamma \end{bmatrix},\quad
\Lambda_2=\beta\gamma\begin{bmatrix}0 & 0 &1\\0 & 0 &0\\
-\cos \varphi & -\sin \varphi &0\end{bmatrix}.
\end{equation}
The kinetic kernel, which can be worked out using equations~\eqref{eq6},
 \eqref{eq19} and \eqref{eq20} turns out to be a constant matrix
\begin{equation}\label{eq21}
 k=\begin{bmatrix} k_1 & k_2\\-k_2 & k_1\end{bmatrix},
\end{equation}
where 
$k_1=\mathbf{\Omega}\cdot\mathbf{I}=\gamma^2\Omega_0I_3$ 
and $k_2=-\mathbf{g}\cdot \mathbf{I}/c =\gamma^2\beta\Omega_0I_1$. 
Here $I_i,(I_i)_{jk}=-\epsilon_{ijk}$, 
is a $3\times 3$ matrix proportional to the operator of infinitesimal rotations about the $x^i$-axis. If in
equation~\eqref{eq16} $a_k=0$ for all integers $k\neq 0$, then $\mathcal{F}=0$ 
and the observed field is 
\begin{equation}\label{eq22} 
F=\frac{1}{2}i\gamma \omega A
\begin{bmatrix} \hat{\mathbf{e}}_{\pm}\\ \hat{\mathbf{b}}_\pm\end{bmatrix}
 \frac{\omega e^{-i\hat{\omega}(\tau-\tau_0)}\mp\Omega_0}{\omega \mp \Omega_0}
+c.\;c.,
\end{equation}
where $\hat{\mathbf{b}}_\pm=\mp i\hat{\mathbf{e}}_\pm$ and
\begin{equation}\label{eq23} 
\hat{\mathbf{e}}_\pm =\frac{1}{\sqrt{2}}
\begin{bmatrix} 1\\ \pm i\gamma^{-1}\\\pm i\beta\end{bmatrix}.
\end{equation} 
This field is only 
due to the kinetic kernel~\cite{8}.
For $\Omega_0/\omega \ll 1$, $F$ involves a small constant term with 
amplitude $\Omega_0/\omega$ and a harmonic term of frequency $\hat{\omega}$, 
as expected.
The predicted constant term is a direct result of nonlocality and
has not yet been experimentally verified; 
however, it may be rather difficult to search for such a term of 
very small amplitude
$\Omega_0/\omega$ in the presence of noise.
\begin{figure}[pht]
\centerline{\epsffile{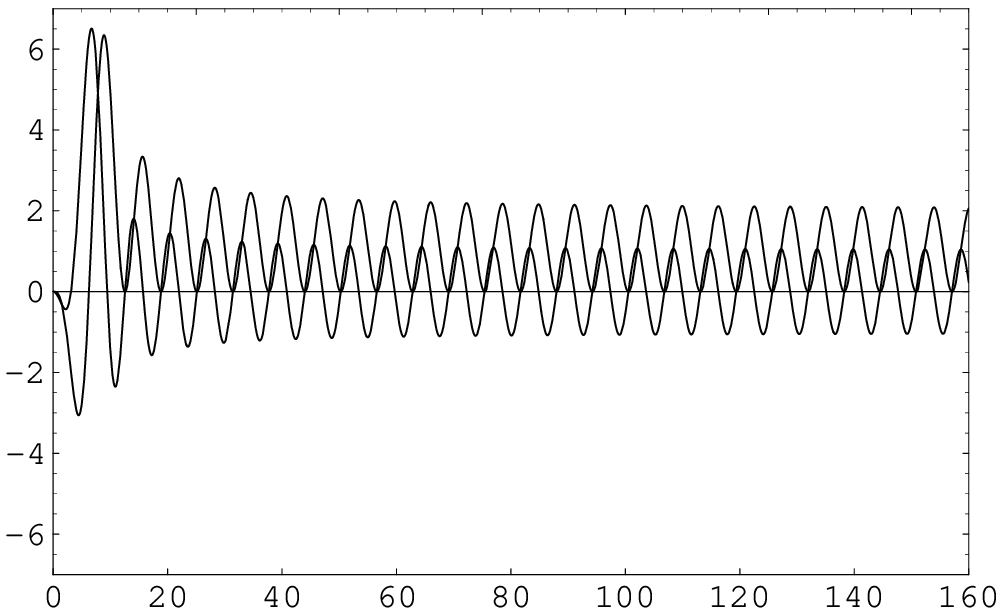}}
\vspace{.2in}
\centerline{\epsffile{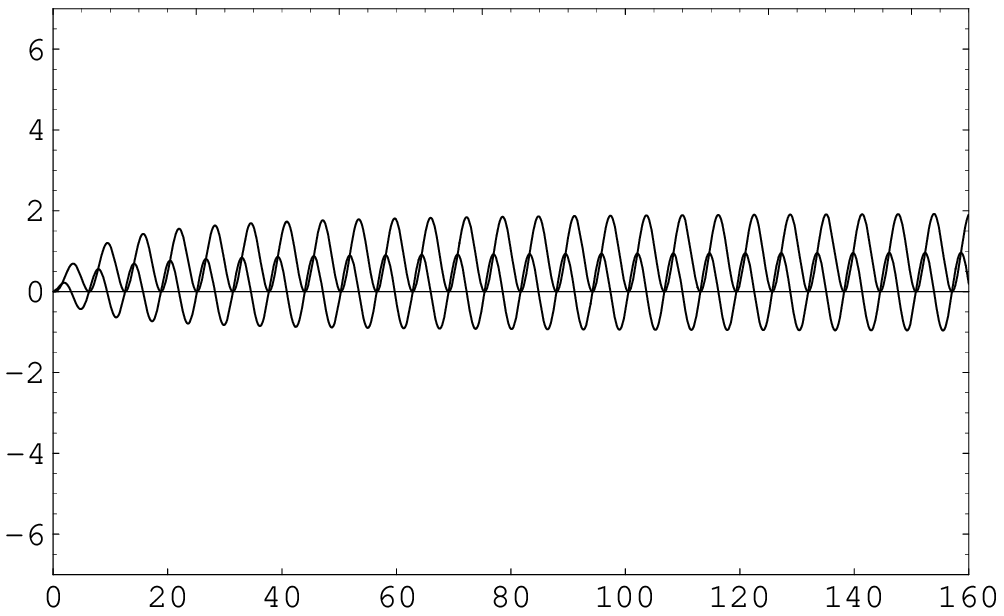}}
\caption{The function ${f}_k(\tau -\tau_0)$ defined by 
equations~\eqref{eq14} and \eqref{eq18} is given, up to 
constant proportionality factors, by $\mathcal{R}_k(x)+i\mathcal{I}_k(x)$
for $x=\gamma \omega (\tau -\tau _0)$. 
Here $k$ is a nonzero integer, $\mathcal{R}_k(x)=(1-2\pi k/x)^{-1}(1-\cos x)$ 
and
$\mathcal{I}_k(x)=(1-2\pi k/x)^{-1}\sin x$. 
These functions are plotted here versus $x$ for $k=1$ in the top panel and
for  $k=-1$ in the bottom panel. 
For $k>0$ we have $\mathcal{R}_k(2\pi k)=0$ and
$\mathcal{I}_k(2\pi k)=2\pi k$;
therefore, the amplitude of the transient could be very large.
The graphs illustrate the fact that the $k$-independent 
steady state is established for $x\gg 2\pi |k|$.}\end{figure}

To find the frequency spectrum of the extra field $\mathcal{F}$, 
let us note that the substitution of equation~\eqref{eq18} in 
equation~\eqref{eq14} results in an
expression for ${f}_k(\tau -\tau_0)$ that exhibits transient as well as 
steady-state behaviors. The nature of the transients is illustrated in 
Figure~1.
For observational purposes, 
we are only interested in the steady-state behavior of ${f}_k$. 
At late times $\tau-\tau_0\gg 2\pi |k|/\omega$, ${f}_k$
approaches a steady state given by
\begin{equation}\label{eq24} 
{f}_k(\tau -\tau_0)\sim \frac{1}{2}\gamma^{-1}A
\begin{bmatrix} \mathbf{e}_\pm \\ \mathbf{b}_\pm \end{bmatrix} (1-e^{-i\gamma
\omega (\tau -\tau _0)})+c.\;c.\,,  
\end{equation}
which is independent of $k$. 
Once the steady state is established, 
${f}_k$ is real and can be expressed as a constant term together with 
a harmonic term of frequency $\gamma\omega$. 
Let us next consider the frequency content of 
$a_k(\tau-\tau_0)$ given by equation~\eqref{eq10}. We note that
\begin{equation}\label{eq25} 
\Lambda \begin{bmatrix} \mathbf{e}_\pm\\\mathbf{b}_\pm\end{bmatrix} 
=\gamma \begin{bmatrix} \hat{\mathbf{e}}_\pm\\
\hat{\mathbf{b}}_\pm \end{bmatrix} e^{\pm i\varphi }.
\end{equation}
We can determine the frequency content of $a_k(\tau-\tau_0)$
by taking derivatives of equation~\eqref{eq25} with respect to $\tau$
and using the fact that for $m>1$
\begin{equation}\label{eq:26}
\frac{d^m\Lambda}{d\tau^m}
\begin{bmatrix} \hat{\mathbf{e}}_\pm\\\hat{\mathbf{b}}_\pm\end{bmatrix}
=(\pm i \gamma \Omega_0)^m\Big\{ 
\begin{bmatrix} \hat{\mathbf{e}}_\pm\\\hat{\mathbf{b}}_\pm\end{bmatrix}
e^{\pm i\varphi}+O(\beta^2)
\Big\},
\end{equation}
where $O(\beta^2)$ indicates terms proportional to either $\cos\varphi$
or $\sin\varphi$ with amplitudes that are smaller than the
corresponding amplitude of the main term by a factor of $\sim \beta^2$.  
Combining these results, we
find that the Fourier content of $\mathcal{F}$ is given in this case as follows:
the term proportional to $Q_k$ consists of a constant plus a harmonic term of
frequency $\gamma \omega$, which disagrees with observation. 
The next term proportional to $C^1_{km}$ consists of harmonic terms with 
frequencies $\hat{\omega}$
and $\gamma \Omega_0$; the latter term is contrary to observation. 
Moreover, the term proportional to $C^n_{km}$ for $n>1$ would contain 
principal harmonics
$\gamma (\omega\mp n\Omega_0)$ and $n\gamma \Omega_0$ that would be 
in contradiction with experimental results as well as terms
whose amplitudes would be smaller by a factor of $\sim \beta^2$. 
We can thus conclude that either
$\mathcal{F}=0$ or that its amplitude is so small as to have escaped 
detection thus far. We may therefore proceed with the tentative assumption 
that
$\mathcal{F}=0$ and the kinetic kernel \eqref{eq6} is unique, while keeping in 
mind the possibility that future experimental data may prompt us to take
$\mathcal{F}$ into account as well. 
This eventuality appears highly unlikely, however, 
from a theoretical standpoint since 
$\hat{\omega}=\gamma (\omega\mp\Omega_0)$ for $\Omega_0\ll \omega$ 
emerges from the simple kinematics of Maxwell's theory \cite{9}.

\section{Discussion}

In this paper we have looked for bounded continuous kernels within the framework of the nonlocal theory of accelerated observers that would be consistent with
Lorentz invariance and satisfy the requirement that a basic radiation field would never stand completely still with respect to an accelerated observer.
Concentrating on electrodynamics and taking into account observational data regarding spin-rotation coupling, we find that the kinetic kernel $K(\tau
,\tau')=k(\tau ')$ given by equation~\eqref{eq6} is the only one consistent with the data thus far. We therefore adopt the kinetic kernel for the nonlocal theory
in general, regardless of the nature of the field.

For the kinetic kernel, the nonlocal contribution to the field has the character of a weighted average such that the weight function is proportional to the
acceleration of the noninertial observer. This result is consistent with the idea put forward by Bohr and Rosenfeld~\cite{15} that the measured field is an average
over a spacetime region. For the case of an accelerated observer, the spacetime region reduces to the past worldline of the observer due to certain basic
limitations on the measurement of distance discussed in \cite{3,4}. Furthermore, Bohr and Rosenfeld~\cite{15} considered only inertial observers for which the
weight function would be unity due to the homogeneity and isotropy of inertial frames of reference. However, the field measurements of a noninertial observer would
be weighted according to its acceleration along its past worldline.

\end{document}